ESA Voyage 2050 White Paper:

# Detecting life outside our solar system with a large high-contrast-imaging mission

*In this white paper, we recommend the European Space Agency plays a proactive role in developing a global collaborative effort to construct a large high-contrast imaging space telescope, e.g. as currently under study by NASA. Such a mission will be needed to characterize a sizable sample of temperate Earth-like planets in the habitable zones of nearby Sun-like stars and to search for extraterrestrial biological activity. We provide an overview of relevant European expertise, and advocate ESA to start a technology development program towards detecting life outside the Solar system.*

**Contact scientist:**

Ignas Snellen, Leiden Observatory, Leiden University, Postbus 9513, 2300 RA Leiden, Netherlands

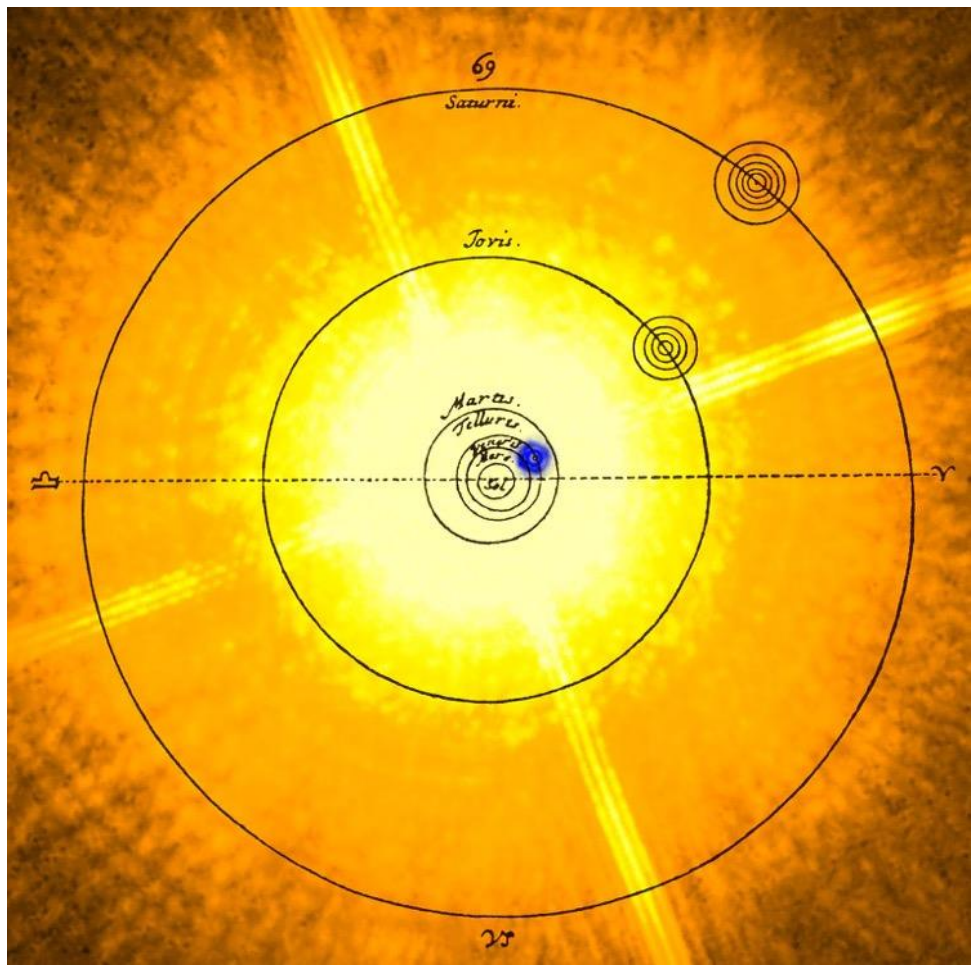

*Image adapted from "Cosmotheoros" (1698) by Christiaan Huygens, one of the first scientific/philosophical treatises about extraterrestrial life. The aim of this white paper is to directly detect signs of life in the light of an Earth-like exoplanet, orbiting a nearby sun-like star, whose light (symbolized here by an HST image of a stellar diffraction halo in the background) is about ten billion times brighter.*

## Executive Summary

Over the last 25 years, thousands of extrasolar planets have been discovered, exhibiting an enormous diversity in mass, size, composition, and orbital properties. Some gas giants are in such tight orbits that they are hotter than most stars in the Milky Way, while others are found at hundreds of astronomical units (AU) distance. Many planets are unlike those in our own solar system, with still to be determined bulk compositions. We *do* know by now that rocky planets are very common, particularly around red dwarf stars, and temperate planets - like Earth - are expected to orbit many of our nearest neighbors. This puts us in a position to start addressing existential questions about the place of humanity and Earth in the Universe. In this white paper, we advocate for a high-contrast-imaging mission to characterize a sizable sample of twin-Earths and search for life.

We can expect that in the next decade, by means of radial velocity, transits, astrometry, and microlensing surveys including ESA's Gaia, CHEOPS and PLATO missions, a general census of the exoplanet population will be largely completed. Atmospheric characterization, however, requires different measurements, since most exoplanets are discovered without identifying photons from the planets themselves. Both transit spectroscopy and high-contrast imaging techniques have already delivered a string of very exciting results, both from space and from the ground. Transit and dayside spectroscopic observations have constrained the climate and temperature structure for a range of hot Jupiters, and have revealed molecular, atomic and ionic absorption features, Rayleigh scattering by hazes, constrained global circulation patterns, and provided evidence for clouds and global winds. High-contrast imaging has provided such evidence for around a dozen directly-imaged massive young gas giants, including their spin rate and cloud-deck variability. Many endeavors are under way to push these measurements to smaller and cooler planets, and the future is bright. In the next decade, the James Webb Space Telescope (JWST), ARIEL, and the Extremely Large Telescopes (ELTs) will come online. They are expected to extend atmospheric characterization to a few of the most favorable temperate rocky planets, such as Proxima b and those in the TRAPPIST-1 system, to address whether such planets have atmospheres, whether they are wet, and potentially habitable.

Characterizing true Earth analogs that are in the Habitable Zones (HZs) of sun-like stars (except for the Alpha Centaurus system) is outside the realm of both the JWST and the ELTs. This requires a large high-contrast-imaging telescope mission, e.g. such as the HabEx and LUVOIR missions currently under study by NASA. We encourage ESA to play a proactive role in building a global collaborative effort to design and construct such a telescope for launch in the 2040s. We provide an overview of relevant European expertise, and advocate ESA to start a technology development program towards detecting life outside the solar system.

*This direct imaging white paper joins two other papers that are advocating interferometry (Quanz et al.) and a starshade (Janson et al.) for the detection of biomarkers.*

# 1. The exoplanet revolution

Since the first discovery of planet-mass bodies orbiting a pulsar (Wolszczan and Frail 1992) and the first gas giant orbiting a solar-type star (Mayor & Queloz 1995), thousands of extrasolar planets have been found. Much of the scientific progress is fueled by enormous improvements in instrumental capabilities and precision, and a steep learning curve on how to optimally utilize both space and ground-based telescopes. The rapid sequence of milestones is amazing: after the discovery of the first transiting planet (Charbonneau et al. 2000; Henry et al. 2000), the first atmospheric feature was detected only two years later (Charbonneau et al. 2002), followed by the first thermal emission measurements (Deming et al. 2005; Charbonneau et al. 2005), the discovery of the first exo-Neptune (Butler et al. 2004) and hot rocky planets (Leger et al. 2009; Batalha et al. 2010), a planet thermal map (Knutson et al. 2007), the first direct image of a planet (Marois et al. 2008; Lagrange et al. 2009), planet spin (Snellen et al. 2014), culminating in the exciting discoveries of the first nearby temperate rocky planets Proxima b and those in the TRAPPIST-1 system (Anglada-Escudé et al. 2016; Gillon et al. 2017) - a true scientific revolution.

Our ultimate goal is to understand the place of Earth and the Solar System in the Universe. How do planets form? What ranges of architectures of planetary systems exist, and how does the solar system fit in? How unique is the Earth as a planetary host of biological activity? To address such existential questions, a broad framework is needed that explains the formation and evolution of planets as a function of mass, composition, birth-orbit and migration history. This will need to be coupled to a thorough understanding of geological processes, atmospheric physics and chemistry for a wide range of planets. Only at that point we will be in a position to identify biomarker gases (e.g. Meadows et al. 2018) as unambiguous proof for biological activity.

In this white paper, we advocate ESA to start a global collaboration to develop and build a large, high-contrast-imaging space telescope, to be launched in 2040. Only a handful of temperate rocky-planets around mid/late M-dwarfs – the low-hanging fruit - will have been characterized by that time. This mission will conduct a large characterization census of twin-Earths around solar-type stars in the solar neighborhood, and the many other planets in these systems, and assess their climate, habitability, and whether they are inhabited. This can only be done with a space telescope. In Section 2 we describe the current status of the field, and in Section 3 the outlook for the 2020-2035 period. Section 4 discusses the science case for this mission and the resulting instrument requirements. Section 5 describes the current HabEx and LUVOIR designs and provides a non-exhaustive list of potential areas for Europe to participate, and a roadmap for a technology development program towards detecting life outside the solar system

This white paper has benefited greatly from excellent recent reviews, notably the *Exoplanet Science Strategy*[1] (2018) and *Astrobiology strategy for the search for life in the universe*[2] (2019), both from the National Academy of Sciences.

## 2. Current status of the field

### 2.1 Finding planets

Most exoplanets have been discovered using the radial velocity (RV) technique, which measures the host star's reflex motion around the system's center of mass, or with the transit method, which targets the regular and temporal dimming of a star caused by a passing planet.

*The RV technique* requires spectrographs with a resolving power of $R \sim 10^5$, and has so far been utilized only on ground-based telescopes. The ultra-stable HARPS spectrograph (Mayor et al. 2003) on the 3.6-m telescope of the European Southern Observatory (ESO) has played a pioneering and leading role. The RV signal induced by an unseen planet scales with the planet mass, square-root of the orbital distance, and the sine of the orbital inclination. Massive planets in close orbits, such as hot Jupiters, are therefore the easiest to detect and produce RV signals of up to 150-200 m/s. An Earth-mass planet in an Earth orbit around a solar-mass star induces an RV signal of 9 cm/s.

*The transit method* relies on wide field CCD photometry to monitor the flux of thousands of stars simultaneously. The transit signal scales with the square of the ratio of planet to star radius, which corresponds to 1% for a Jupiter-size planet transiting a solar-type star and around 0.01% for an Earth-size planet. The duration of a transit is typically one to a few hours for a close-in planet, to half a day for a planet in an Earth-size orbit. Transit surveys are strongly biased to planets with short orbital periods, because those in wider orbits have a lower probability to transit, and if they do, produce transits less frequently. Ground based transit surveys have resulted in hundreds of hot and warm gas-giant planets (e.g. Alonso et al. 2004; Street et al. 2002; Bakos et al. 2002). The CoRoT and NASA Kepler space mission have probed down to rocky planet sizes, albeit mostly for relatively faint and distant stars. In particular, Kepler has resulted in thousands of discoveries, reaching down to Earth-size planets in Earth-size orbits around solar type stars (e.g. Petigura et al. 2013).

---

[1] NAS Exoplanet Science Strategy (2018) available at http://nap.edu/25187
[2] NAS Astrobiology strategy for the search for life in the universe (2019); http://nap.edu/25252

_High-contrast imaging_ is currently exclusively performed by ground-based instrumentation, and relies on adaptive optics techniques that can correct the stellar wavefront from atmospheric turbulence, and coronagraphy to enhance the achieved contrast by blocking the starlight as much as possible, while letting the planet light pass through. Planets that are relatively bright and far enough from their host star are the most accessible, and current discoveries are mostly limited to young (10-100 Myr) self-luminous gas giant planets orbiting at tens of AU or more. Particularly remarkable examples are Beta Pictoris b in a nearly edge-on 10 AU orbit (Lagrange et al. 2019), the four-planet system of HR8799 (Marois et al. 2011), and the two planets in the PDS 70 system which still contains a gas & dust disk apparently feeding the (proto-)planets (Muller et al. 2018; Keppler et al. 2018; Haffert et al. 2019a) - see Figure 1. The few dozen gas giants and sub-stellar companions known so far are seen at contrasts of typically $10^{4-5}$. At mature ages of a billion years or more, these planets will have faded by several orders of magnitude below current detection limits.

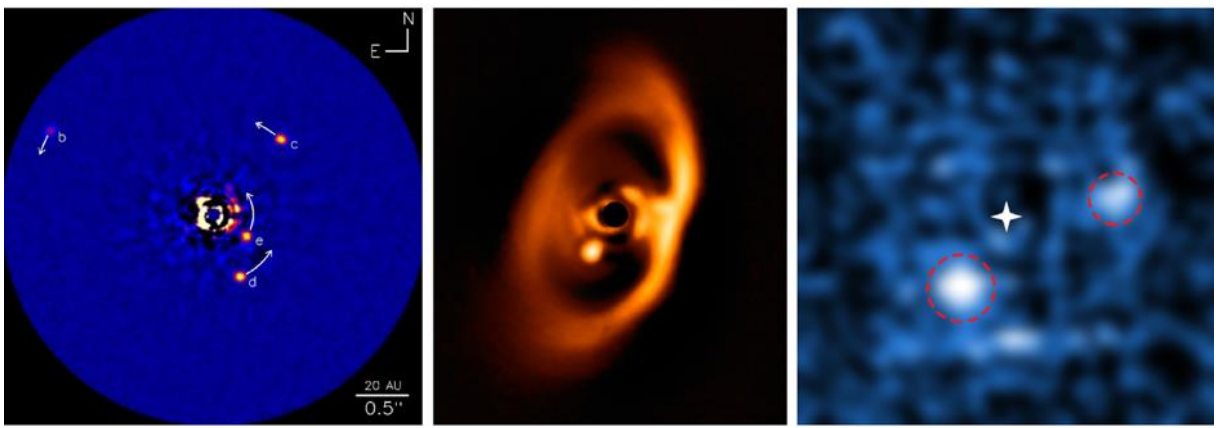

**Figure 1:** *The two directly imaged systems with multiple planets so far, with HR 8799 (left) and PDS 70 with SPHERE (middle; Keppler et al. 2018) and in Hα (Haffert et al. 2019a). Image credits: NRC-HIA/C.Marois/W.M.Keck Observatory; ESO/Müller et al; ESO/S. Haffert, Leiden Observatory.*

_Microlensing_ provides additional information on the architecture of planetary systems as the observational biases are very different from the other techniques. A foreground star with accompanying planet can act as a gravitational lens of a background star and significantly boost its flux. Although this technique does not provide precise measurements of a planet's mass and orbit, a large ensemble can provide statistics on planets in relatively wide (4-10 AU) orbits that are otherwise inaccessible with the other methods. Microlensing surveys are currently conducted by ground-based telescopes mostly monitoring very wide star fields towards the bulge. As soon as a week- to month-long stellar microlensing event is identified, the target is then intensively monitored to identify short-duration amplifications due to planets in the system.

Finding planets through _astrometry_, which measures the reflex motion of the star in the plane of the sky, has held tremendous promise from the earliest days in the exoplanet field. Only now --- with ESA's Gaia mission --- is this technique expected to deliver (Perryman et al. 2014). Interestingly, the observational bias on orbital radius is opposite to those for the RV and transit techniques, generating a larger signal for planets in wider orbits (with a limit set by the total mission duration). Astrometric signals are inversely proportional with distance, making the technique most sensitive to the most nearby stars.

Red dwarf stars are particularly interesting for finding planets. Since stellar masses are up to an order of magnitude smaller than that of solar-mass stars the RV signals are correspondingly larger - and since stellar radii are up to a factor ten smaller, the corresponding transit signal for a given radius planet can be up to two orders of magnitude larger. Their luminosities can be a thousand times lower meaning that temperate planets can be found in smaller orbits – making even small, rocky planets accessible through both the transit and the RV technique. This is called the _red dwarf opportunity_.

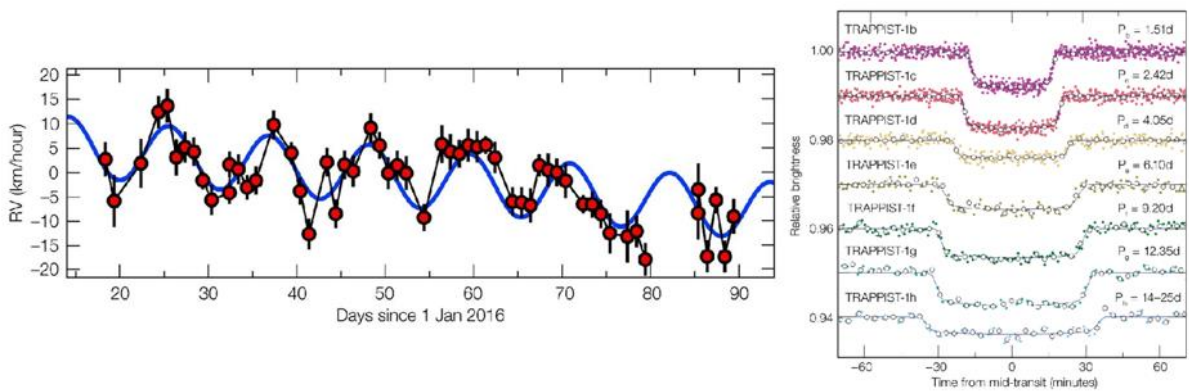

*Figure 2:* [Left] Discovery RV measurements from HARPS showing the signal from Proxima Cen b. [Right] The transit signals of the seven Earth-size planets in the TRAPPIST-1 system. Credits: ESO/G. Anglada-Escudé & ESO/M. Gillon et al.

Several, relatively small, ground based projects are currently targeting transits of mid-M dwarfs (e.g. MEarth[3]) and late M-dwarfs (TRAPPIST; SPECULOOS[4]; Gillon et al. 2017), and significant observing time is being invested in RV surveys targeting the nearest red dwarf population (e.g. Pale Red Dot[5]). The most thrilling discoveries are those of Proxima b, a planet with a minimum mass of 1.3 $M_{Earth}$ in an 11-day orbit (meaning it receives about 70% of stellar energy compared to the Earth - see Figure 2; Anglada-Escudé et al. 2016), and the seven Earth-size planets in the TRAPPIST-1 system (Gillon et al. 2017) of which at least three have equilibrium temperatures that indicate they could have liquid water on their surfaces - see Figure 2.

## 2.2 Planet demographics

A key finding of exoplanet research so far is that planets are very common and diverse – something that was completely unknown 25 years ago. Although the results of early RV and transit surveys were dominated by detections of hot Jupiters, this is largely caused by observational biases. These planets are actually relatively rare – occurring around typically 1% of FGK stars, dropping to even lower fractions around the lowest mass stars (Mayor et al. 2011). They are likely the product of extreme orbital migration, either due to early interactions with the protoplanetary disk and/or other objects in the system. The occurrence rate of gas giant planets out to a few AU is about 10-15%, with many of them in eccentric orbits. Such planets are found to be significantly more common around higher metallicity stars (Santos et al. 2004) – a correlation not found for lower mass planets.

Kepler has shown that Neptune size planets in close-in orbits (<<1 AU) are common (Howard et al. 2012), and that about half of the stars have planets in size between the Earth and Neptune in orbits smaller than Mercury's. The distribution in planet radius and orbital distance shows a valley of low number density separating gas-rich planets from possible rock-dominated planets, a separation which could be caused by processes of mass loss due to photo-evaporation (Fulton et al. 2017; Van Eylen et al. 2018).

---

[3] https://www.cfa.harvard.edu/MEarth/Welcome.html

[4] https://www.eso.org/public/chile/teles-instr/paranal-observatory/speculoos/

[5] https://palereddot.org

While the Kepler mission has just fallen short of an accurate determination of the occurrence rate of temperate Earth-size planet around FGK stars, it is clear that they are very abundant (>10% - Burke et al. 2015). They may even be more ubiquitous around M-dwarfs, with estimates of one HZ planet with 1-1.5 $R_{Earth}$ for every four M-dwarfs (Dressing & Charbonneau 2015).

The relatively few detections of young gas giant planets with direct imaging suggest that their occurrence rate at wide (>30 AU) orbital separations is 0.01-1% (e.g. Bowler et al. 2016), including significant uncertainties due to the steep age-luminosity and mass-luminosity relations for gas giant planets, and uncertainty in the initial entropy of formation. Current statistics from microlensing surveys are the most detailed for M-dwarf systems. They show that Jupiter and Neptune mass planets are less common than around solar-type stars, but there exists a significant population of cold super-Earths in relatively wide orbits (e.g. Clanton & Gaudi 2016).

## 2.3 Planet characterization – bulk compositions and atmospheres

Radial velocity measurements of a transiting exoplanet system reveal the mass of the planet without inclination ambiguity, which in combination with the planet radius will provide an estimate of the planet's mean density. Although this has so far been only possible for planets in relatively short orbits, it is clear that _bulk compositions_ can vary wildly from planet to planet – even for planets with the same mass or radius. The density of Jupiter-like planets show a wide range of variation, probably because of differing rocky core fractions and due to various levels of inflation from incident host star radiation. First indications are that the core masses of gas giants in wider orbits correlate with the metallicity of their host stars (Thorngren et al. 2016).

Lower mass planets are found to have increasingly higher densities, consisting of varying combinations of iron, rock, ice, and gas – which are not straightforward to solve due to degeneracies in internal structure models. It is expected that such degeneracies can at least be partly resolved by future atmospheric observations. Current statistics are highly biased towards planets in the shortest orbits. They imply that there is an overlap between terrestrial super-Earths and gas-rich mini-Neptunes with the same mass, but significantly different radii. Below a size of ~1.6 $R_{Earth}$, planets are mostly found to be rocky with compositions consistent with that of the Earth (e.g. Dressing et al. 2015).

Most exoplanets are discovered without identifying a single photon from the planets themselves. For _atmospheric characterization_, planet light needs to be separated from that of the star. There are generally two families of techniques that can accomplish this. The first relies on temporal variations of the planet and starlight – these include transmission and secondary eclipse spectroscopy and planet phase curve observations. Since they mostly (but not exclusively) rely on transiting systems, and transit probabilities are lower for planets in increasingly wider orbits, such observations are performed on stars that are relatively distant and faint. The second family of techniques relies on angularly separating the planet light from that of the star, i.e. high contrast imaging – any direct image of a planet automatically contains information on the planet's atmospheric properties.

During a transit, starlight is blocked by the planet, but also a small fraction of starlight filters through the planet's atmosphere where it is partially absorbed and scattered by atoms, molecules, and/or cloud and haze particles. When the planet itself is eclipsed, comparison with observations just before and after the eclipse can reveal reflected light and direct thermal emission from the planet, which has also an imprint of possible atmospheric absorption and scattering processes. Observations during other parts of the orbit can reveal variations in emissions from the planet when different fractions of the dayside and nightside hemispheres of the planets come into view.

Gas giant planets in close-in orbits provide the most accessible signals for these time-differential techniques. Their high temperatures result in the most favorable planet/star contrasts for eclipse spectroscopy and, in combination with their $H_2$ atmospheres, the largest transmission signals. Observations have mainly been performed with the Hubble and Spitzer Space Telescopes, but also from the ground – in particular using spectroscopy at very high resolving powers of R~100,000 which helps to remove stellar and telluric contamination (e.g. Snellen et al. 2010). Observations of hot Jupiters have shown the presence of Rayleigh scattering from high altitude hazes (Sing et al. 2016), and atmospheric absorption from atomic gases such as sodium and iron, ionized iron and titanium (e.g. Hoeijmakers et al. 2018), and molecules such as water, carbon monoxide, and titanium oxide (Deming et al. 2013; Nugroho et al. 2017). Some planets show significant atmospheric losses through their evaporating exosphere which contains atomic hydrogen and helium (e.g. Ehrenreich et al. 2015; Spake et al. 2018). Eclipse and phase curve observations constrain the vertical and longitudinal temperature structure of these planets, including the presence of thermal inversions. Their day-to-nightside temperature variations provide clues on global circulation processes and overall climate. High-altitude winds are also directly probed with high-dispersion spectroscopy. The extension of these techniques to smaller and cooler planets have revealed the presence of high altitude clouds that mask possible molecular absorption features (Kreidberg et al. 2014). This is a major concern, in particular for transmission spectroscopy, which may be mitigated by observing in the mid-infrared, e.g. with MIRI/JWST where the clouds are expected to become transparent.

For high contrast imaging, young gas giants (1-50 Myr) in wide orbits are best accessible due to their relatively large angular separation from their host stars and favorable contrast due to residual heat from their formation. As for eclipse observations, this constrains the temperature-pressure profiles of these planets and molecular content such as for water, carbon monoxide and methane (e.g. Chilcote et al. 2017). Planet spin rotation has been determined from line broadening (Snellen et al. 2014) and the presence of a varying cloud cover has been established through planet flux variations.

More recently the first exciting exoplanet results have been presented using optical-NIR interferometry using the GRAVITY instrument of ESO's VLTI platform (Lacour et al. 2019). We like to note that mid-infrared space-based interferometry is the subject of a white paper lead by Sascha Quanz.

# 3. Outlook for the 2020 – 2035 period
## 3.1 A census of the planet population in the Milky Way

As this white paper advocates for the development of a large space telescope for high contrast imaging of nearby planetary systems, we will focus here on the question what will be learned about the nearby planet population in the next 15 years. The NASA TESS mission[6] is currently conducting an almost complete all sky transit survey. It will finish its Southern + Northern hemisphere surveys before the end of 2020, and will likely continue to operate for several years after – conducting an ecliptic survey, and/or revisiting the previous survey areas for one or multiple visits. With its larger telescope, the ESA CHEOPS mission[7] (launch 2019) will perform crucial follow up studies of the most interesting systems, while the ESA PLATO mission[8] will combine unprecedented photometric precision with a very wide field of view to determine the percentage of solar type stars with temperate Earth-size planets. There will also still be an important role for ground-based transit surveys – such as SPECULOOS and ExTrA[9] (which also uses spectral information) – which will focus on planets transiting the smallest M-dwarf stars. Such stars are optically too faint and dispersed across the sky to be covered by the wide angle space-based surveys. With all these new and ongoing missions and surveys it is to be expected that nearly all nearby transiting systems will have been discovered by 2035. However, the closest Earth transiting a solar-type star will not be quite 'nearby'. Since the transit probability is on the order of 0.5%, there is a non-transiting Earth-twin to be expected about six times closer to us --- making it uncertain whether such transiting Earth-twins will be preferable targets for high-contrast imaging.

---

[6] https://www.nasa.gov/tess-transiting-exoplanet-survey-satellite
[7] http://sci.esa.int/cosmic-vision/53541-cheops-definition-study-report-red-book/
[8] http://sci.esa.int/plato/59252-plato-definition-study-report-red-book/#
[9] https://www.eso.org/public/teles-instr/lasilla/extra/

Ground-based RV discoveries will further accelerate with the many existing and upcoming surveys, such as with HARPS+NIRPS, HARPS-North, ESPRESSO, and CARMENES (to name the main European-led projects) obtaining large fractions of their telescope time. Since the high speed at which discoveries are strung together, it is to be expected that most planets around nearby M-dwarfs will have been found by 2035, with temperate planets already found around Proxima (1.4 pc; Anglada-Escudé et al. 2016), Barnard's star (1.8 pc; Ribas et al. 2018), Lalande 21185 (2.5 pc; Diaz et al. 2019), Wolf 359 (2.4 pc; Tuomi et al. 2019) and Teegarden's star (3.8 pc; Zechmeister et al. 2019). While instrumental stability is expected to approach the level required to detect Earth twins around solar type stars (9 cm/s), astrophysical noise from granulation, pulsations, rotation modulated star spots, and magnetic cycles will make it difficult to reach this level in practice. Several avenues are being explored to separate out true planet-induced RV signals from stellar noise. Acquiring spectra at particularly high signal-to-noise can reveal subsets of spectral lines that react differently to stellar effects (e.g. Dumusque et al. 2018), and lines in the near infrared are expected to be less sensitive to activity. The Terra Hunting Experiment[10] will monitor fifty nearby solar type stars every night for ten years to optimize the observational window function – hoping to reach Earth twins in that way. However, also such experiment will select the quietest stars, and not necessarily the nearest. It is therefore not clear yet whether the Earth twins around the nearest sun-like stars will be known by 2035. Measuring the masses of such planets through RV will be significantly easier afterwards, when the orbital periods are known through high-contrast imaging with the mission proposed here. A long-term high-precision RV monitoring campaign for the most nearby solar-type stars is therefore highly recommendable.

The full data release of the ESA Gaia mission will contain thousands of gas giant planets in relatively wide orbits around nearby stars. It will determine the architecture of the outer parts of the planetary systems around virtually all of our neighbors. NASA's WFIRST mission will conduct microlensing surveys that will provide the population statistics of outer planetary systems on a wider scale (and to a much lower mass) – generally from systems at hundreds to thousands of parsecs.

Assuming that the high contrast imaging mode of WFIRST will only be a technological demonstrator, high contrast imaging will, also in the 2020-2035 timeframe, be advanced from the ground only. The ELTs (2025-2030) will not only play an exciting role in characterization of the nearest exoplanet systems (see below), but also in discovering planets. At 10 microns, the METIS instrument[11] on the European ELT can detect an Earth twin in the Alpha Cen system within one night – a goal that is already being pursued using the VLT through the Breakthrough Initiative program (Kasper et al. 2017).

Since the diameter of the European ELT is about five times larger than that of the VLT, the collecting area is 25 times larger, while its angular resolution (theoretically) is 5 times better, resulting in 25 times less background and stellar noise at the planet's location. This will make the ELT (in the extreme adaptive optics regime) at least a few hundred to a thousand times faster than the VLT, and also many new targets will become accessible at smaller angles only accessible by the ELT. While current young stellar system surveys are barely scratching the surface of unveiling the (proto-)planet population, this will be rapidly advanced with the ELTs – probing significantly lower masses and older systems. In combination with the revolutionary work by ALMA, it can be expected that the general paradigm of planet formation will be set in the 2035 timeframe. Of course, certain specific aspects will require significant more work, similar to today's understanding of stellar and galaxy evolution theory.

---

[10] http://www.terrahunting.org/index.html
[11] http://metis.strw.leidenuniv.nl

## 3.2 Exoplanet characterization

Exoplanet characterization from space will be dominated by the JWST (2021) and ESA's ARIEL mission[12] (2028). Compared to the Hubble and Spitzer Space Telescopes, the JWST will be a giant leap forward in transmission and secondary eclipse spectroscopy, utilizing a collecting area that is respectively 6.3 and 44 times larger. For the first time a wide spectral range from red-optical to 20 microns will be covered (e.g. Greene et al. 2016). ARIEL will consist of a significantly smaller telescope, but it is one that will be fully dedicated to exoplanet characterization, concentrating on a large sample of hot and warm gas giants and smaller planets. It will form the basis of mapping atmospheric physics and chemistry for a wide range of planets in different environments and with different evolutionary histories.

How far can the JWST be pushed remains to be seen, but it is clear that great strides forward will be made. It is planned to measure the secondary eclipse depths of the TRAPPIST-1b planet (GTO time) at 12 and 15 microns, determining whether this planet has an atmosphere, and whether it contains carbon dioxide (e.g. Lustig-Yaeger et al. 2019). Already these measurements will have an enormous impact, revealing to what extent rocky planets around late M-dwarfs can retain their atmosphere, and can be wet and habitable. There are three orders of magnitude of science between what we study now - mainly hot Jupiter atmospheres - and our goal of studying the habitats of temperate rocky planets. The JWST will cover a significant fraction of that ground, but will not be able to characterize Earth twins.

Three ELT projects are on track to see first light in the 2025-2030 time frame: the European ELT (39m), the Giant Magellan Telescope (GMT; 25m), and the Thirty Meter Telescope (TMT; 30m). Concentrating on the European perspective, the ELT will host as first-light instrument METIS, which is capable of detecting Proxima b by combining high-contrast imaging with high-dispersion spectroscopy (e.g. Lovis et al. 2017) – focusing on spectral signatures of water, carbon dioxide, methane, and even heavy water – HDO (Molliere & Snellen 2019). Other very nearby temperate planets will also become accessible with HIRES – the optical-NIR high-dispersion spectrograph, expected as a second-generation ELT instrument – and capable of detecting molecular oxygen. The ELT will be pushed to its limits when such instruments are combined with extreme adaptive optics providing superb wavefront correction (e.g. PCS/EPICS). While this is clearly technically challenging, it would allow measurements of albedo, overall climate, and molecular absorption such as from water, carbon dioxide, methane, and oxygen (Figure 3) for a handful of temperate rocky planets around the nearest M-dwarfs.

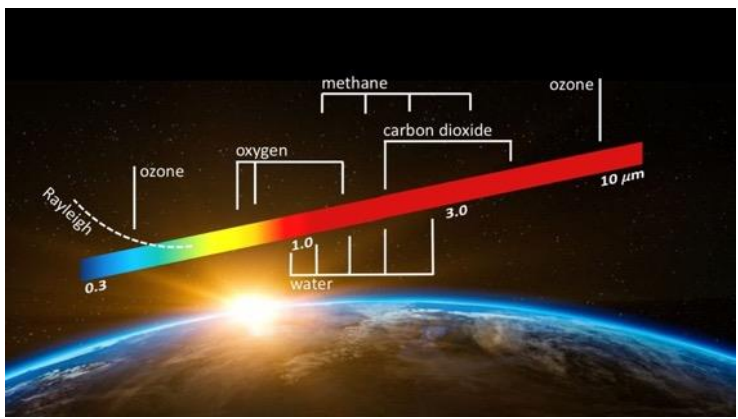

*Figure 3:* Schematic overview of important molecular features from the UV, optical, to the infrared.

---

[12] http://sci.esa.int/ariel

# 4. Exoplanet & Astrobiology Science 2035 – 2050
*The case for a large high-contrast-imaging space telescope*

As argued above, by 2035 we will have discovered most temperate planets around nearby (<5 pc) M-dwarfs. The nearest and most interesting transiting systems, both around M-dwarfs and solar-type stars, will also have been found. The nearest Earth twins (non-transiting around solar type stars) are only accessible with RV, and will remain a challenge. The large high contrast imaging space telescope, as advocated here, may have to determine their orbits first (which is not an issue, because they are very common), before (pre-existing) RV data can reveal their masses.

A handful of temperate, terrestrial planets that orbit the nearest M-dwarfs will have been characterized in detail with the ELTs. We will likely know their main atmospheric constituencies, have constrained their climates and assessed their habitability. Veritable Earth-twins, those that orbit solar-type stars, will be out of reach of ground-based characterization due to their extreme planet/star contrasts. They form, together with the many other planets in their systems, the hunting ground for a large high-contrast-imaging space telescope. Below we discuss the main science questions and topics that such space telescope will address. We have finally reached a level of technical ability such that we can fulfill our long-standing desire to study Earth-twins and search for extraterrestrial life.

## 4.1 Climate & Habitability

As exoplanet science is an extreme form of remote sensing, a biosignature is defined as an effect of biological activity on its environment that is detectable at interstellar distances. This is expected to require a surface biosphere needing the support of a surface ocean (in contrast to e.g. a subsurface biosphere; Des Marais et al., 2002; Schwieterman et al., 2018). This drives the definition of the HZ as the region around a star where a planet with an Earth atmosphere can maintain liquid surface water (Kasting et al. 1993; Kopparapu et al. 2018). Supported by the fact that all life on Earth is carbon based, carbon is likely the only element in the periodic table that can form the basis of a sufficiently wide variety of complex molecules that can undergo complex chemistry, and thus can form living organisms. However, it is not clear exactly how life formed on Earth, nor can we yet produce life from inorganic material in the laboratory. We *do* know that life on Earth formed relatively quickly, implying that life will develop if the circumstances are right.

Several factors are expected to make a planet habitable, or at least impact its habitability, such as:

- <u>A liquid ocean on its surface</u>. This will depend on the balance between how much water is delivered to the planet during formation and its subsequent water loss.
- <u>A secondary atmosphere</u>, assuming that $H_2$-dominated atmospheres are not habitable (Pierrehumbert & Gaidos 2011)
- <u>Geological activity such as volcanism</u>, to counteract atmospheric loss and protect against climate change (Walker et al., 1981).
- <u>A global magnetic field</u>, which depends on the composition and evolution of the planet (e.g. Driscoll and Bercovici, 2013).

The internal energy budget of the planet is likely to be important, which is dependent on its thermal history since formation and on possible tidal interactions with the host star and/or other planets in the system. It will be crucial to understand targets in the context of their host star and planetary system environments.

It is currently not clear whether M dwarfs, the targets of pre-2035 ground based studies, can harbor habitable planets. Firstly, they have an up to 1-billion year pre-main-sequence phase during which the stars are significantly more luminous. This is expected to drive strong atmospheric and ocean loss. Secondly, M-dwarfs can exhibit strong chromospheric activity, producing flares that can briefly increase the stellar UV flux by 1-2 orders of magnitude or more, potentially accompanied by charged particle ejecta. It is not clear how long-term exposure to such flares could affect a possible atmosphere and biosphere.

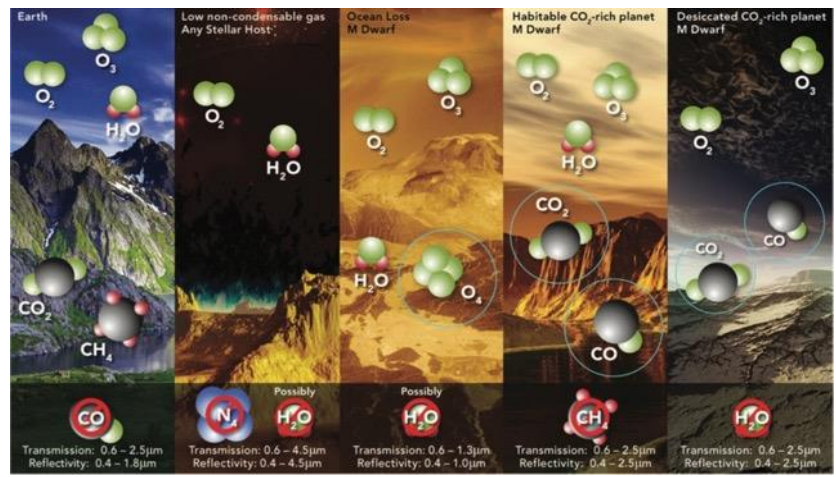

*Figure 4:* From Meadows et al. (2018): Overview of possible false positive mechanisms for molecular oxygen. Detection of circled and forbidden molecules will separate out biotic and abiotic processes (credit; V. Meadows).

In addition, temperate planets around late M-dwarfs are expected to be tidally locked, significantly affecting their global circulation flows and climate. E.g. it is not clear to what extent volatiles, like water, could be frozen out on an eternal nightside (e.g. Ribas et al. 2016; Luger & Barnes 2015; Garcia-Sage 2017; Rugheimer et al. 2015). It is evident that true twin-Earths around solar-type stars will need to be studied to obtain a complete picture of the possible habitability and biological activity in the solar neighborhood. This can only be done with a space telescope, needed to reach the extreme star/planet contrasts required.

The prime focus of the large high-contrast-imaging space telescope is to:

1. survey a sufficient number of stars to find ten(s) of exo-Earths (<1.5 $R_{Earth}$) in the HZs of sun-like stars in the solar neighborhood;
2. determine their planetary system architectures, understand the evolutionary history of the system, and current interplay between the planets;
3. determine and derive whether the planets have atmospheres, and if so their main atmospheric constituencies - for all accessible planets in the systems;
4. determine the overall climates of these planets, assess the diversity in planetary systems and understand their physical and chemical atmospheric processes, and possible geological influences;
5. find out whether the planets contain water, and therefore could be habitable.
6. Identify biomarker gases and determine whether they originate from biotic or abiotic processes.

## 4.2 Biomarker gases and the search for biological activity

It is clear that to unambiguously identify biomarker gases as sign posts of biological activity, we need a deep understanding of both planet formation and evolutionary histories, and of the physical/chemical atmospheric and geological processes on the planet. This is why it is key to not only study potential habitable and biologically active planets in isolation, but to understand the system in context and compare their properties with those of inhabitable and inhabited planets.

Important recent work has evaluated molecular oxygen ($O_2$) as a potential biosignature – originally identified as a promising pathway to find extraterrestrial life in exoplanet atmospheres by Lovelock (1965). It is currently the most detectable signal of life in the Earth's atmosphere (20% by volume), created as a waste product of oxygenic photosynthesis. Photochemistry of $O_2$ also leads to the production of ozone.

New studies have identified different abiotic ways to form $O_2$, which can be recognized using their environmental context (e.g Meadows et al. 2018; Figure 4). Most of these abiotic scenarios involve very different surface conditions and evolutionary histories as Earth, such as photochemical processes for planets with high carbon dioxide concentrations, or planets with significant loss of water through photo-dissociation. The latter can even result in oxygen dominated atmospheres with multiple bars of $O_2$. These abiotic processes can all be distinguished in different ways. They either require a very dry planet such that measuring water features will be sufficient, or by the detection of methane (which excludes an oxygen dominated atmosphere), or the absence of very deep and broad features of carbon dioxide which would disprove it as a dominant atmospheric gas. Inclusion of these environmental contexts has greatly strengthened the use of molecular oxygen as a reliable signature of biological activity. In addition, the first steps are being taken to identify and interpret other potential biomarker gases, such as methyl chloride and dimethyl sulfide, including those that permit detection of non-Earth-like metabolisms.

## 4.3 Science requirements for a >2035 high-contrast imaging space mission

**Direct imaging with a large space telescope is the only way to obtain spectra of the atmospheres of rocky planets in the HZs of solar type stars in a systematic way, because ground based instrumentation is unable to reach the extreme contrasts required.**

Light emerging from the surface of the planet contains two components, that of intrinsic thermal emission and scattered (reflected) starlight. For temperate rocky planets, the thermal component significantly improves the planet/star contrast beyond 5 microns in wavelength, but since telescope angular resolution is inversely proportional to the wavelength, it would require an unfeasibly large telescope to directly image a sample of such planets in the mid-infrared. This can only be reached with a space-based nulling-interferometer, the subject of a different white paper (led by Sascha Quanz). In this white paper, we explore and advocate the only plausible alternative: to concentrate on starlight that is scattered by the planet's atmosphere, either through Rayleigh or Mie scattering, or reflection from its surface. This starlight will be altered by atomic and molecular absorption on the way in and out of the planet's atmosphere, which therefore provides a measure of its constituents, physical and chemical (and possibly biological) processes, and climate.

A twin Earth around a solar-type star will have a contrast in reflected light (depending on its orbital phase and inclination) of at least a few billion, requiring a contrast of $10^{-10}$. To study a sizable sample of 10 to 20 Earth twins in the HZs of FGK stars requires a sample of 75-150 solar-type stars, corresponding to a maximum distance of 10 to 13 pc, which requires an inner working angle (IWA) of 100 to 80 milliarcseconds (mas). Of course, for any telescope this will depend on the observed wavelength (see below).

*Requirement 1: The instrument needs to reach a contrast of $10^{-10}$ at an IWA of 80 – 100 mas.*

In the planet spectra, we will be most interested in scattering processes and molecular absorption features. Rayleigh scattering will be strongest in the UV and optical-blue. Several water vapor bands are present from 0.7 micron onwards and are gradually stronger towards longer wavelength (see Figure 3). The strongest feature of molecular oxygen is at 0.76 micron, with other bands at 0.69 and 1.27 micron, while ozone shows a cutoff at 0.33 with a shallow feature at 0.55 micron. Methane has the strongest features at 2.2 and 3.2 micron, with a weaker band at 1.6 micron, while carbon dioxide has strong bands around 2.1, 2.8 and 4.3 micron.

***Requirement 2: The wavelength range needs to cover the optical and near-infrared, with a goal to also include the UV and mid-infrared up to 5 micron.***

The observed planet spectra will strongly depend on the reflective and scattering properties of the atmosphere as a function of altitude, and the volume mixing ratios of the spectroscopically active molecules. Since several features can overlap, it will be important to have a sufficiently high spectral resolution and sufficient wavelength range to identify multiple absorption bands of the same molecule.

***Requirement 3: The spectral resolution needs to be sufficient to uniquely identify molecular absorption.***

The mission will need to be able to survey the planetary systems of 75 to 150 stars, and obtain their spectra at a sufficiently high signal-to-noise.

***Requirement 4: The light collecting power and mission life time of the space telescope should be such that spectra of 75 – 150 planetary systems can be obtained.***

In the next section we will present and discuss the main characteristics of such a space mission. Currently, NASA is performing two concept studies of mission that could fulfill the requirements listed above. One is the Habitable Exoplanet Observatory (HabEx), a 4m monolithic telescope operating between 0.2 and 1.8 micron with an external starshade. The other is the Large UV/Optical/IR surveyor (LUVOIR), an 8 to 15m general purpose telescope covering 0.2 to 2.5 micron. These are both very exciting studies that investigate different ways of how the requirements above can be met, that will identify and map technical challenges, and will provide first cost estimates of different mission architectures.

Requirements 1, 2 and 4 on contrast, IWA, wavelength range, and sample size, are closely coupled to each other. The IWA governs to what distance a twin-Earth planet can still be separated from its parent star, and is set by the Rayleigh criterion ($\lambda/D$) and/or the size and distance of the external occulter or starshade in the case of a HabEx-type design. At a distance of 10 (13) parsec this corresponds to a maximum IWA of 100 (77) mas. Assuming that ultimately a space telescope plus internal coronagraph will reach its extreme contrast limits down to $2.5\lambda/D$, observations at a wavelength of 0.5, 1.0 and 2 micron will require telescopes of at least 3m (4m), 6m (8m), and 12m (16m) in size respectively (see Figure 5). External occulters have a different dependency, and the one currently under study for HabEx is envisaged to have an IWA of 40 and 100 mas at UV/Optical and infrared wavelengths respectively. This is the main reason why NASA is studying these different mission concepts of a 4m (HabEx + starshade), 8m (LUVOIR, architecture B) and 15m (LUVOIR, architecture A).

In addition to spectroscopy, also the measurement of the polarization of the scattered/reflected light from an exoplanet enables the characterization of its atmosphere and surface, particularly in the combined method of spectropolarimetry. Such observations provide unique and unambiguous constraints on water clouds (Karalidi et al. 2012a) and surface oceans, while they also enhance the contrast of biomarker molecule spectral lines (Stam 2008). Moreover, both spectroscopy and polarimetry allow us to distinguish the light from exoplanets from residual starlight and light from exozodiacal dust (Perrin et al. 2015). Finally, circular spectropolarimetry constitutes an ultimate probe of biological activity, as it is sensitive to the homochirality of biotic molecules (Patty et al. 2019), although it may require an even larger space telescope or a very favorable target to detect such signatures.

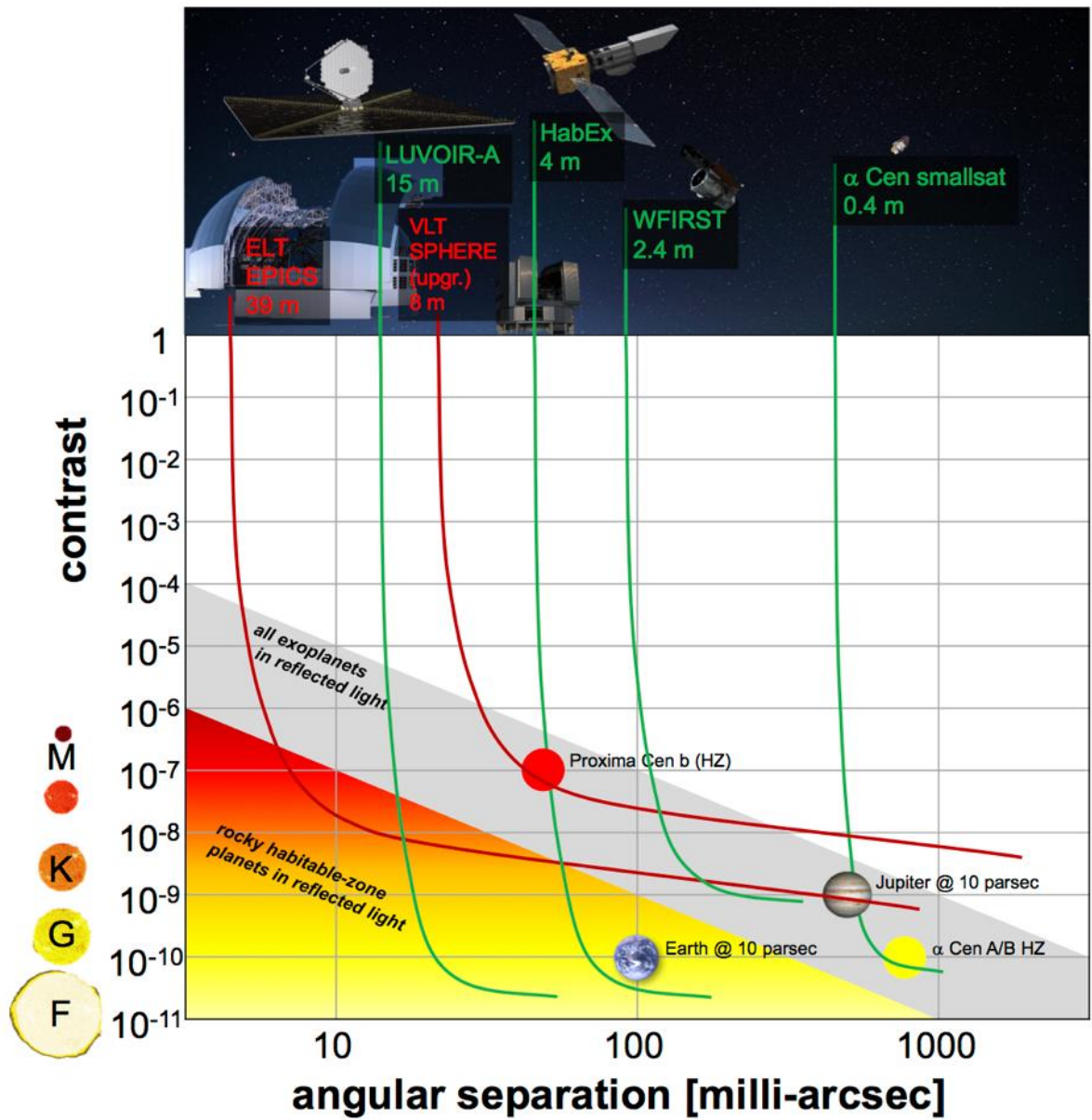

*Figure 5: Approximate projected reflected-light total contrast versus angular separation for future ground-based and space-based telescopes. HZ planets lie in the lower left hand corner, and this region is only accessible for M stars from the ground for ELTs and solar-type stars are reachable with smaller diameter telescopes in space. These curves are approximate and for illustration purposes only.*

After the conclusion of the first LUVOIR & HabEx mission design studies in the coming year, the process of the US decadal survey, and the successful launch of the JWST, it will be perfect timing for Europe to start to seriously contribute to the discussions of what type of mission design would be most effective and cost efficient in addressing the enigmatic and exciting questions about the place of Earth and humanity in the Universe, and to play an important and guiding role in realizing this exciting goal.

# 5. HabEx, LUVOIR, and potential areas for ESA contributions

## 5.1 Challenges of high-contrast Imaging in Space

High Contrast Imaging (HCI) techniques aim to optimize the detection of the signal from scarce photons scattered or radiated by an exoplanet by removing the diffracted light halo of its parent star. To reach extreme levels of contrast, several complementary optical and data-reduction techniques need to work in very close harmony. See Ruane et al., Jovanonic et al., and Snik et al. (2018) for recent reviews of HCI techniques.

HCI is becoming a mature technique on ground-based telescopes, with still rapidly expanding technological and scientific capabilities for dedicated instruments like VLT/SPHERE (Beuzit et al. 2019), Gemini/GPI (Macintosh et al. 2018), Subaru/SCExAO (Lozi et al. 2018), and Magallan/MagAO(-X) (Males et al. 2018). HCI from the ground is fundamentally limited by the Earth's atmosphere, as even the most advanced Adaptive Optics (AO) systems are not fast enough to provide perfect correction of atmospheric turbulence for even the brightest stars (that also feed the wavefront sensor). The "raw" contrast is therefore limited to $\sim 10^{-5}$ at small angular separations, as rapidly varying atmospheric speckles and slowly varying instrumental speckles still create a halo of starlight on the detector. On the ground, differential techniques that distinguish light from exoplanets from residual starlight (e.g. based on their different spatial, spectral, polarization and coherent properties) are therefore vital to directly detect and characterize even the brightest exoplanets.

In space, the stable environment and long exposure times enable a raw contrast as deep as $\sim 10^{-10}$ (Figure 5). The level of control over the optics in a space-based HCI instrument is therefore of a completely different magnitude than for ground-based instruments, although the time-scales involved are slower. The necessary ingredients for a HCI space mission include:

- **A large, optimized telescope** to efficiently collect the sparse photons from the faint exoplanet targets, and to deliver a diffraction-limited performance with an angular resolving power of $\sim \lambda/D$. We adopt the visible band as the shortest wavelength range, as it contains essential biomarkers, and technology is mature for HCI (in contrast to the UV range), and therefore the telescope diameter D needs to be as large as possible to enable observations at the smallest possible angular separations.
- **Adaptive optics** with two deformable mirrors to actively control both the phase and the amplitude of the wavefront propagating through the instrument at extreme precision to ensure stability of the point-spread function (PSF) at a level of $10^{-10}$.
- **Coronagraphy** to selectively suppress the on-axis stellar PSF and its diffraction structure, whilst transmitting the exoplanet PSF for the smallest possible off-axis angular separation. A coronagraph is a trade-off between planet throughput, IWA and their tolerance to errors from telescope optics and their stability. Internal coronagraphs suppress starlight after it has entered the telescope, while an external occulter (starshade) at a large distance from the telescope can guarantee large contrast without stringent requirements on the optics.
- **Wavefront / electric field sensing techniques** to control the AO system (preferably with signals measured at the actual science focal plane) to dig "dark holes" in the residual stellar halo, in tandem with the coronagraph architecture.
- **Spectroscopy** to characterize the light captured for an exoplanet. Typically, extreme contrasts can only be attained in narrow spectral bandwidths (~20%), so the spectroscopic implementation depends on the AO+coronagraph implementation. In addition, single-mode fibers can be used to both feed a spectrograph, and also contribute to the overall contrast (Por & Haffert, Haffert et al. 2019b).

- **Polarization techniques** to mitigate the detrimental polarization-dependent aberrations of any optical system (Breckinridge et al. 2015, Schmid et al. 2018), to enable the implementation of broadband liquid-crystal-based coronagraphs (Mawet et al. 2010, Snik et al. 2012, 2014), and to enable a measurement of exoplanet polarization to both enhance the contrast as well as provide complementary diagnostics to spectroscopy.
- **Sensitive detectors** to perform imaging/spectroscopy/polarimetry, without adding noise sources more severe than the photon noise. Specific detector architectures can also be used for focal-plane wavefront sensing: the Self-Coherent Camera (SCC; Baudoz et al. 2006), or the noise-less, wavelength resolving Microwave Kinetic Inductance detector (MKID; Bueno et al. 2018).
- **Integrated system design approaches** to make sure that all listed subsystems actually perform better than just the sum of their parts, and that optimal solutions are chosen based on the overall technical/scientific performance.
- **Data-reduction techniques** to optimally extract the exoplanetary signals to determine their orbits, phase angles, size/albedo, spectral and polarization signals.
- **Mission planning/yield optimization** to observe the maximum number of targets given mission lifetime (and, e.g., starshade repointing), including the required revisits, and long exposures for characterization once a promising candidate has been pinned down.
- **Data-analysis methods** to interpret the acquired spectral and other data in terms of, e.g., atmospheric parameters and constituents.

## 5.2 Mission architectures

This white paper highlights the science and science requirements for a future HCI mission to detect biomarkers in the light from Earth-like exoplanets orbiting sun-like stars. We advocate that ESA and the European exoplanet community take a leading role towards the realization of such a mission and all its aspects. It is to be expected that this endeavor that we will carry out on behalf of all of humanity will also be a truly global enterprise with several international partners. We do not state any hard preferences regarding technology or collaborative model. In this section, we present the current status of the NASA-led design studies HabEx and LUVOIR, that could fulfill our shared scientific goals, and are open to collaboration with external partners such as ESA (but are not guaranteed to fly, with a first decision pending the US AstroDecadal2020 process). But, most of all, the designs for HabEx and LUVOIR provide an excellent review of the state-of-the-art.

### 5.2.1 Current baseline design of the HabEx mission

The Habitable Exoplanet Observatory (HabEx[13]; Gaudi et al. 2028) is one of the mission concepts currently under study by NASA. Its baseline design consists of a monolithic, off-axis 4-m telescope in an L2 orbit, which is diffraction limited at 0.4 micron. Currently, it has two ways to suppress starlight, with an internal coronagraph and with an external starshade. Both are envisioned to have dedicated instruments for imaging and spectroscopy. In the current study, a mission lifetime of five years is assumed.

---
[13] https://www.jpl.nasa.gov/habex/

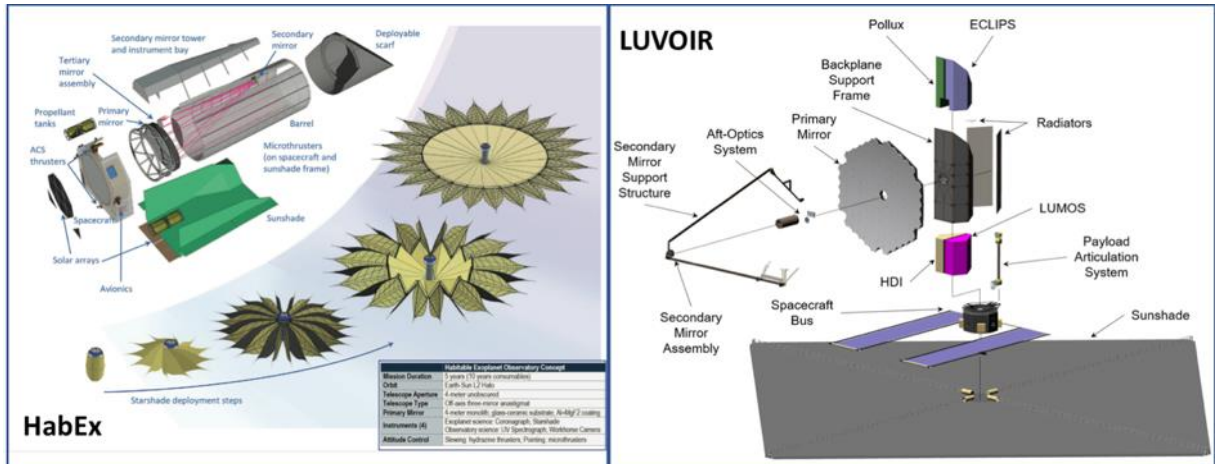

*Figure 6:* *The HabEx and LUVOIR concept designs (credit: LUVOIR Final, and HabEx Interim Reports).*

The off-axis monolithic mirror avoids the challenges encountered by centrally obscured or segmented mirrors to obtain coronagraphic performances sufficiently high close to the star, and with a sufficiently high throughput. HabEx has a vortex coronagraph in its baseline design, which works well in the presence of low-order wavefront (and polarization) aberrations, meaning less stringent requirements on thermal and mechanical stability of the telescope. It produces a raw contrast of $10^{-10}$ at an IWA of 62 mas at 0.5 micron with a 20% bandwidth. In the study, the coronagraph can be used with several cameras and integral field spectrograph covering 0.45 to 0.67 micron, 0.67 to 1 micron, and an IR imaging spectrograph covering 0.95 to 1.8 microns.

The baseline study also envisions a starshade of 52-m diameter that would fly in formation with the telescope at typically 100 thousand kilometers distance. The advantage of blocking the starlight before it enters the telescope are mainly a high throughput, small IWA and an outer working angle only set by the detector size. In this case, an IWA of 60 mas at 1 micron is expected to be achieved. The distance between the starshade and the telescope can also be varied to optimize for observations at shorter or longer wavelengths. The starshade instrument is currently baselined as having three channels, covering 0.2 to 0.45 micron with a grism, 0.45 to 1 micron with an integral field spectrograph, and 1 to 1.8 micron also with an IFS.

### 5.2.2 Current baseline design of the LUVOIR mission

The Large Ultraviolet / Optical / Infrared Surveyor, LUVOIR[14] (the LUVOIR team, 2018) is another mission concept currently under study at NASA, with baseline architectures A and B consisting of a 15-m on-axis and 8-m off-axis telescope respectively. They have segmented mirrors to fit in a single launch vehicle. It is envisioned to feature a coronagraph with imaging camera and integral field spectrograph covering 0.2 to 2 micron. A serviceable design is an integral part of its architecture. Both the A and B architectures are designed to operate at L2 for a 5 year prime mission lifetime. LUVOIR leans on heritage from JWST, such as deployable telescopes and segmented wavefront control. The telescopes are diffraction limited at 0.5 micron. An effort is underway to design a high-performance coronagraph coupled to a segmented, obscured telescope. In addition, technologies need to be developed for stabilization of the wavefront coming from the telescope involving low-order wavefront sensors and deformable mirrors. The highest contrast measured to date at testbeds at JPL is 10$^{-9}$. Ultimately, contrasts of 10$^{-10}$ are envisioned at 3.5 λ/D over a total passband of 0.2 to 2 micron. Current baseline of the spectrograph has a resolving power of R=140 in the visual, and R=70 and 200 in the near-infrared.

---

[14] https://asd.gsfc.nasa.gov/luvoir/

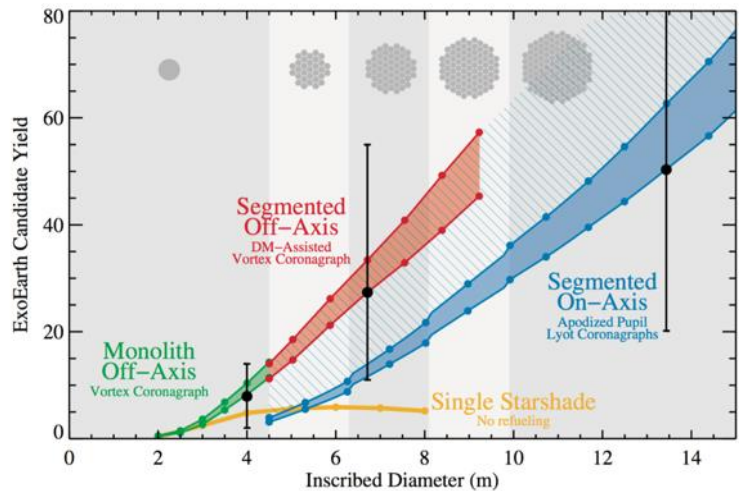

*Figure 8:* *From Stark et al. 2019: Study of ExoEarth yield as a function of telescope aperture and off-axis or on-axis architectures. This is normalized for a mission of 2 years, and assuming η $_{Earth}$ = 0.24. Note that the twin-Earth yield does not scale linearly with mission lifetime, because the best accessible systems get observed first (Credit: Stark et al. 2019).*

## 5.3 Potential Earth-twin yield of different mission designs

The HabEx and LUVOIR A and B architectures have on key aspects different designs, such as on-axis versus off-axis, monolithic versus segmented mirrors, and differences in aperture size. Stark et al. (2019) have carried out a preliminary study to assess the impact of these design choices on the expected exo-Earth yields. They consider four telescope designs, one with an external occulter (a starshade) and three designs with internal coronagraphs. They conclude that, currently, an off-axis design significantly improves the final expected exo-Earth yield. The coronagraph design tradeoff with a secondary obscuration is sensitive to the IWA and planet flux throughput, which directly impacts the maximum distance of the stellar system that can be reached for a given telescope diameter D. For off-axis telescopes there seems to be little difference between monolithic or segmented mirrors.

The results from Stark et al. (2019) are summarized in Figure 8. The yield of 8 twin-Earths for the HabEx mission is for a two-year mission lifetime, assuming η$_{Earth}$ = 0.24. Note that this does not scale linearly with survey time (which may be 5 or even 10 years) because it requires more distant and difficult targets to be imaged.

## 5.4 A technological and scientific development path for Europe

The scientific goals of this white paper are extremely ambitious, but address possibly the most exciting question humanity has been asking itself: Are we alone? To achieve the required performance and collaboration for a space mission as outlined above that will answer this question by 2050, ESA and the European scientific community need to embark as soon as possible on a "voyage" that has several intermediate destinations:

- **Start a European development program for technology validation and cost/risk reduction.** In Sect. 5.5 we identify an initial selection of specific European expertise to invest in.
- Establish a European strategy and collaborative model regarding **testbeds for technology demonstration** for HCI in space. Such laboratory experiments are crucial to validate new technologies, techniques and system level approaches, and drive their performance to the extremely challenging requirements introduced in Sect. 4.3.
- **Increase the Technology Readiness Level** (TRL) of specific technologies or subsystems with experiments on stratospheric balloons and smallsats. Such small missions could even resolve the habitable zone around our neighbor solar-type stars α Cen A/B (see Fig 5).
- Identify possible opportunities for **full science+technology demonstrator missions**, that perform important observations of gas giant or super-Earth-like exoplanets. Currently, the CGI instrument on WFIRST is the only space mission which aims to detect and characterize planets in reflected light. The science yields of WFIRST-CGI (Mennesson et al. 2018) indicates that dozens of giant planets, known from RV detection, can be detected in photometry, while perhaps a single one can be observed in spectroscopy. The Earth twin science case is far enough from the expected WFIRST yield to motivate an intermediate step with a compelling and unique science case, unfeasible from the ground before 2050. In the context of an M class proposal to Cosmic Vision (2010), the scientific potential was investigated of a small (1.5m) off-axis optimized telescope to study giant planet atmospheres (SPICES, Boccaletti et al. 2012; Maire et al. 2012). A similar mission with a larger size telescope will definitely make a compelling science case as an intermediate step towards the large HCI space telescope as described in this white paper. Furthermore, a dedicated starshade mission in combination with existing telescopes (ground or space-based) would enable direct observations of exoplanets at extremely high contrast (Janson 2007).
- Perform **benchmark observations of the Earth as an exoplanet**, e.g. by a dedicated small instrument on/around the moon (Karalidi et al. 2012b), or from a cubesat or the ISS, can further inform design decisions for a large HCI instrument to observe Earth-twins, and guide the development of data-analysis techniques and improve models of the Earth's atmosphere and surface, which are essential for detecting life on an exo-Earth, and for understanding the effects of our life on Earth (e.g. climate change, biodiversity loss, etc.).
- To achieve a high-confidence detection of life on an exoplanet, we need to **form an interdisciplinary community** of astronomers, geophysicists, technical physicists (optics, detectors, etc. specialists), biologists, ecologists, climate scientists, etc. to be able to fully interpret the observations in all its aspects. Moreover, we need to collaborate with researchers in the fields of the humanities, philosophers, artists, and of course society at large to give meaning to such a detection of life elsewhere in the universe.

## 5.5 Identification of areas for Europe to participate

The table below indicates the current expertise from European researchers. This is meant to be indicative and not an exhaustive list of technology areas we recommend ESA to further invest in.

| Large space telescopes | ● Euclid high precision optical telescope (Wachter and Markovic 2018; Wallner et al. 2017)<br>● ESA deployable mirror development (Marchi et al. 2017) |
|---|---|
| Adaptive Optics | ● SPHERE extreme AO system: (Fusco et al. 2006; Petit et al. 2014; Beuzit et al. 2019)<br>● The très haute dynamique bench (THD; Galicher et al. 2014, Baudoz et al. 2018)<br>● Using 2 DMs for phase/amplitude control (Mazoyer et al. 2017)<br>● ESA active optics developments (Hallibert & Marchi 2016; Laslandes et al. 2017)<br>● Deformable Mirror development (Charlton et al. 2014) |
| Coronagraphy | ● 4QPM coronagraph for JWST (Boccaletti et al. 2004, Baudoz et al. 2006b)<br>● APLC coronagraph (N'Diaye et al. 2015, 2016a)<br>● Coronagraph optimization (Carlotti et al. 2014)<br>● AGPM/Vortex coronagraphs (Forsberg and Karlsson 2013; Delacroix et al. 2013)<br>● Advanced liquid crystal coronagraphs (Snik et al. 2012, Doelman et al. 2017; Por et al. 2018; Snik et al. 2018) |
| Wavefront / electric field sensing | ● SCC (Baudoz et al. 2006a; Galicher et al. 2008)<br>● ZELDA Zernike WFS (N'Diaye et al. 2013, 2016b, Vigan et al. 2019)<br>● vector-Zernike WFS (Doelman et al. 2019)<br>● Pyramid WFS (Ragazzoni et al. 2017)<br>● Segmented space telescope phasing (JWST+LUVOIR; Leboulleux et al. 2018)<br>● Speckle nulling (Martinache et al. 2014)<br>● COFFEE phase diversity (Paul et al. 2014)<br>● Holographic Modal WFS (Wilby et al. 2017), incl EFC (Por & Keller 2016)<br>● Phase-Sorting Interferometry (Codona and Kenworthy 2013)<br>● Asymmetric Pupil-WFS (Martinache et al. 2013)<br>● vAPP fpWFS (Bos et al. 2019)<br>● QACITS algorithm (Huby et al. 2015) |
| Spectroscopy | ● High-contrast imaging + High-resolution spectroscopy: Snellen et al. (2014, 2015), Vigan et al. (2018)<br>● SPHERE microlens-based IFS (Claudi et al. 2006)<br>● Slicer IFS: SINFONI (Thatte et al. 1998), HARMONI (Thatte et al. 2014)<br>● SCAR coronagraph + single-mode fiber spectrograph (Haffert et al. 2019; Por & Haffert 2019b) |
| Polarization techniques | ● Polarization-based 4QPM and VVC coronagraph (Mawet et al. 2006)<br>● Liquid-crystal coronagraphy + polarization filtering (Snik et al. 2014b)<br>● SPHERE-ZIMPOL (Schmid et al. 2018)<br>● Advanced polarimetric techniques: Snik & Keller 2013; Snik et al. (2014a) |
| Detectors | ● MKID detector development for visible light: Baselmans et al. (2017), Bueno et al. (2018) |
| Astrophotonics | ● Photonic reformatting - NAIR (Harris et al. 2018)<br>● 3D printed microlenses on single mode fibre IFUs (Dietrich et al. 2017; Haffert et al. 2019c) |
| System design | ● SPICES HCI space telescope concept (Boccaletti et al. 2012) |
| Data-reduction techniques | ● Spectral Differential Imaging (Claudi et al. 2008, Vigan et al. 2010)<br>● Principal Component Analysis (Amara & Quanz, 2012)<br>● ANDROMEDA (Cantalloube et al. 2015)<br>● ALICE (Choquet et al. 2014) |

**A census of support from the Exoplanet Community**

Although white papers for the ESA Voyage are mainly meant to argue why a specific scientific theme should have priority in the Voyage 2050 planning cycle, and are not proposals for specific missions, we did feel it would be useful and relevant to gauge support for a European contribution to a large high-contrast-imaging space mission among the Exoplanet Community. We therefore contacted key scientists in the field - covering a wide range of exoplanet disciplines. The reactions were overwhelmingly supportive, and a list is provided here.

**Belgium**
Michaël Gillon (University of Liège)
Anne-Lise Maire (University of Liège)

**Denmark**
Lars Buchhave (DTU Copenhagen)
Simon Albrecht (Aarhus University)

**France**
Anne-Marie Lagrange (Grenoble University)
Mamadou N'Diaye (Obs. Cote d'Azur)
Arthur Vigan (LAM Marseille)
Pierre Baudoz (Observatoire de Paris)
Elsa Huby (Observatoire de Paris)
Anthony Boccaletti (Observatoire de Paris)
Franck Selsis (University of Bordeaux)
David Mouillet (Grenoble University)
Jean-Luc Beuzit (LAM Marseille)

**Germany**
Ludmila Carone (MPIA Heidelberg)
Thomas Henning (MPIA Heidelberg)
Markus Kasper (ESO)
Oliver Krause (MPIA Heidelberg)
Heike Rauer (DLR Berlin)
Julien Milli (ESO)
John Lee Grenfell (DLR Berlin)
Roy van Boekel (MPIA Heidelberg)
Ralf Launhardt (MPIA Heidelberg)

**Ireland**
Ernst de Mooij (DCU Dublin)

**The Netherlands**
Frans Snik (Leiden University)
Matthew Kenworthy (Leiden University)
Jos de Boer (Leiden University)
Ignas Snellen (Leiden University)
Yamila Miguel (Leiden University)
Jayne Birkby (University of Amsterdam)
Michiel Min (SRON)
Daphne Stam (Technical University Delft)
Pieter de Visser (SRON)
Jean-Michel Désert (University of Amsterdam)
Christoph Keller (Leiden University)

**Italy**
Alessandro Sozzetti (INAF – Torino)
Giuseppina Micela (INAF – Palermo)
Raffaele Gratton (INAF – Padova)
Silvano Desidera (INAF – Padova)
Riccardo Claudi (INAF – Padova)
Valentina D'Orazi (INAF – Padova)
Isabella Pagano (INAF – Catania)
Giampaolo Piotto (Padova University)

**Spain**
Manuel Lopez-Puertas (IAA Granada)
Ignasi Ribas (ICE Barcelona)
Enric Palle (IAC Tenerife)

**Sweden**
Markus Janson (University of Stockholm)

**Switzerland**
Willy Benz (University of Bern)
Kevin Heng (University of Bern)
Brice-Olivier Demory (University of Bern)

**United Kingdom**
Didier Queloz (University of Cambridge)
Matteo Brogi (University of Warwick)
Mark Claire (University of St. Andrews)
Beth Biller (University of Edinburgh)
Christiane Helling (University of St. Andrews)
Nikku Madhusudhan (Univ. of Cambridge)
Guillem Anglada-Escudé (QMU London)
Nathan Mayne (University of Exeter)
Tim Lenton (University of Exeter)
Isabelle Baraffe (University of Exeter)
Sasha Hinkley (University of Exeter)

**United States of America**
Laura Kreidberg (Harvard University)
Olivier Guyon (University of Arizona; HabEx)
Bertrand Mennesson (NASA JPL; HabEx)
Christopher Stark (STSCI; HabEx/LUVOIR)
Victoria Meadows (Washington; LUVOIR)
Scott Gaudi (Ohio State University; HabEx)
Garreth Ruane (Caltech)

**Proposing team**

| | |
|---|---|
| Ignas Snellen, | Leiden Observatory, The Netherlands |
| Simon Albrecht, | Aarhus University, Denmark |
| Willy Benz, | University of Bern, Switzerland |
| Anthony Boccaletti, | Observatoire de Paris, France |
| Jos de Boer, | Leiden Observatory, The Netherlands |
| Matteo Brogi, | University of Warwick, United Kingdom |
| Lars Buchhave, | DTU Copenhagen, Denmark |
| Riccardo Claudi, | INAF Padova, Italy |
| Rafaelle Gratton, | INAF Padova, Italy |
| Kevin Heng, | University of Bern, Switzerland |
| Thomas Henning, | MPIA Heidelberg, Germany |
| Elsa Huby, | Observatoire de Paris, France |
| Markus Jason, | University of Stockholm, Sweden |
| Markus Kasper, | ESO, Germany |
| Matthew Kenworthy, | Leiden Observatory, The Netherlands |
| Anne-Marie Lagrange, | Grenoble University, France |
| Giuseppina Micela, | INAF Palermo, Italy |
| Yamila Miguel, | Leiden Observatory, The Netherlands |
| Michiel Min, | SRON, The Netherlands |
| Ernst de Mooij, | DCU Dublin, Ireland |
| Mamadou N'Diaye, | Observatoire de la Côte d'Azur, France |
| Isabella Pagano, | INAF Catania, Italy |
| Enric Palle, | IAC Tenerife, Spain |
| Didier Queloz, | University of Cambridge, United Kingdom |
| Heike Rauer, | DLR Berlin, Germany |
| Ignasi Ribas, | ICE Barcelona, Spain |
| Frans Snik, | Leiden Observatory, The Netherlands |
| Alessandro Sozzetti, | INAF Torino, Italy |
| Daphne Stam | TU Delft, The Netherlands |
| Arthur Vigan, | LAM Marseille, France |